# Detection and Analysis of Solar Eclipse

S. S. Sarrvesh, S. Pradeep sundar, I. Kenny Jackson, P. Kannan
Department of Electronics and communication Engineering, Panimalar Engineering College,
Chennai, India.
sarrfriend@gmail.com
sundar.pradeep@gmail.com
kennyjackson@gmail.com
pkannan2003@yahoo.com

*Abstract* – **We propose an algorithm that can be used by amateur astronomers to analyze the images acquired during solar eclipses. The proposed algorithm analyzes the image, detects the eclipse and produces results for parameters like magnitude of eclipse, eclipse obscuration and the approximate distance between the Earth and the Moon.**

## I. INTRODUCTION

Solar eclipses are spectacular celestial events that attract people from both the scientific and the non-scientific community. Professional astronomers, from the scientific community, use various high-end equipments to study the eclipse and its influence, like the temperature variations and gravitational anomalies. For the amateur astronomers, the high-end equipments are usually unavailable or unaffordable: all they can do during an eclipse is just to acquire the images using simple telescopes.

For the sake of amateur astronomers, we propose an algorithm that, when implemented as hardware, can analyze the images taken during a solar eclipse and determine the magnitude of eclipse, eclipse obscuration and the approximate distance between the earth and the moon.

## II. PROPOSED ALGORITHM

For the algorithm to produce the results, it has to extract four values from the image – center of the Sun, center of the Moon, radius of the Sun and radius of the Moon. Note that these parameters are in accordance with the image and are not the actual center and radius. Furthermore, the algorithm does not depend on the method of image acquisition and simply assumes that the input image is a binary version of the image acquired during an eclipse. (The algorithm also assumes that the image was acquired on a cloudless environment. Presence of cloud makes the algorithm to produce false results.)

### A. Sun Tracer Block

The first step is to detect the edges in the binary image using any of the standard edge detection schemes like Prewitt or Sobel schemes. The purpose of the sun tracer block is to obtain two points on the circumference of the Sun so that those points can be used to obtain the center and the radius of the Sun. A simple way to obtain the points is to trace the image pixel-by-pixel for the first high intensity value from both the ends of the image – top end and the bottom end. The pixels that correspond to the first high intensity value (bit level 1) in both the tracing are taken to be the points on the circumference of the Sun.

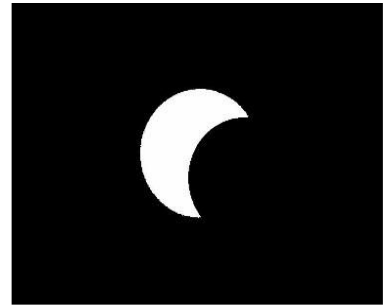

Figure 1. Sample input binary image.

### B. Evaluation of Sun's center and radius

From the two points determined in the above step, the center of the Sun is obtained. With the two traced points as reference, the center point between these two points is obtained. The perpendicular line passing through the newly obtained point is drawn using simple equations from analytical geometry. The radius of the Sun can be calculated by using the distance between the mid-point of the chord and one of its end points and the distance between the mid-point of the chord and the point where the perpendicular line meets the circumference of the circle. The formula to calculate the radius is

$$r = \frac{c\_s^2 + s\_s^2}{2*\sqrt{s\_s}} \quad \text{\_\_\_\_\_\_\_\_\_\_\_\_\_\_\_\_} (1)$$

where c_s is the distance between the mid-point of the chord and one of its end points while s_s is the distance between the mid-point of the chord and the point where the perpendicular line meets the circumference of the circle.

With the radius of the Sun and any two points on the circumference of the circle, the center of the Sun can be determined using the fact that the center is equidistant from both the points.

## C. Extracting the Moon's curve from the Image

Now, with the parameters of the Sun – radius and its center – are calculated, the same parameters of that of the Moon are obtained by performing the same operations on the curve of the Moon. To do these operations, the curve that represents the partial surface of the Moon must be extracted.

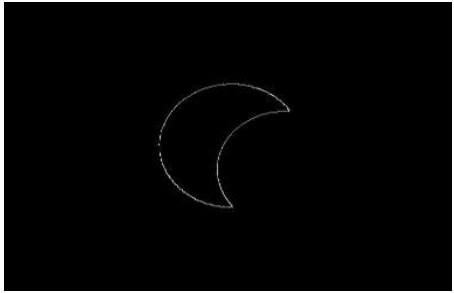

Figure 2. Edge detected version of the input image.

A simple mask that works based on the data obtained from the previous section is used. From the radius and the center of the Sun obtained from the previous section (1), a simple mask that covers about 95% of the Sun's disk is designed and the brighter region inside the disk is extracted. (Or, the Moon's curve is obtained by making all the pixels outside the mask "dark" or bit level-0.) This brighter region represents the partial surface of the Moon. The accuracy of the extracted surface can be improved by adjusting the size of the mask used. As the radius of the mask approaches the radius of the Sun, the accuracy of the extracted surface approaches maximum.

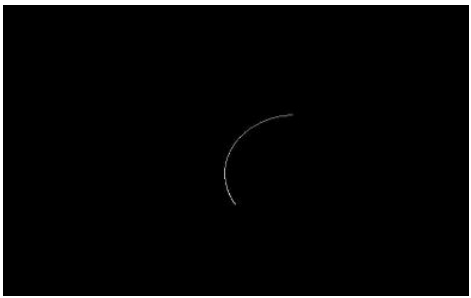

Figure 3. Extracted curve of the Moon.

## D. Moon Tracer Block

The Moon tracer block works exactly the same way as the Sun tracer block. The function of the tracer block is to "trace" to points on the curve of the Moon. As explained earlier, a simple way to obtain the points is to trace the image pixel-by-pixel for the first high intensity value from both the ends of the image – top end and the bottom end. The pixels that correspond to the first high intensity value (bit level 1) in both the tracing are taken to be the points on the circumference of the Moon.

With the edge of the Moon, the center and the apparent radius of the Moon can be calculated by the same process used for the Sun. While calculating the center and the radius of the image, if the algorithm does not detect any curve, it can be inferred that the Moon is not present in the image. This implies that the image acquired contains the Sun alone and thus solar eclipse has not occurred.

## E. Calculating the magnitude of eclipse

The magnitude of eclipse is defined as the ratio of the apparent length of the diameter of the Sun obscured by the Moon to the apparent diameter of the Sun. Solar eclipse can be geometrically visualized as the intersection of two circles. To calculate the magnitude of eclipse, consider the image displayed below that shows the intersection of two circles.

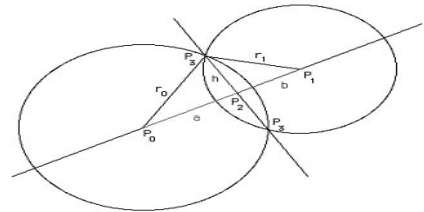

Figure 4. Intersection of two circles used in the calculation of the magnitude of eclipse.

According to the definition, the magnitude of eclipse is given by the formula

$$Eclipse\_Mag = \frac{Len\_Obs}{2 * Sun\_radius} \quad --(2)$$

where Len_Obs is the length of the Sun's diameter obscured by the eclipsing object. Len_Obs is given by the formula

$$Len\_Obs = (Sun\_Radius - a) + (Moon\_Radius - b) \quad --(3)$$

where 'a' is the distance between the points $P_0$ and $P_2$ and 'b' is the distance between the points $P_1$ and $P_2$.

## F. Calculating eclipse obscuration

Eclipse obscuration is defined as the fraction of the Sun's area occulted by the Moon. Mathematically speaking, this can be visualized as the area of the region produced by the intersection of two circles. Now, consider two circles of radii R and r and centered at (0, 0) and (d, 0) intersect in a region shaped like an asymmetric lens.

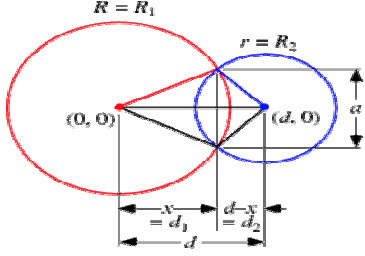

Figure 5. Calculation of eclipse obscuration

The formula used to calculate the eclipse obscuration is

$$Eclipse\_Obs = \frac{OverlappedArea}{AreaOfSun} \quad (4)$$

where OverlappedArea is the area of the asymmetric lens shown in figure 5. The formula to calculate the overlapped area is

$$r^2 Cos^{-1}\left[\frac{d^2+r^2+R^2}{2dr}\right] + R^2 Cos^{-1}\left[\frac{d^2+R^2-r^2}{2dR}\right] - 0.5\sqrt{(-d+r+R)(d+r-R)(d-r+R)(d+r+R)} \quad (5)$$

During a solar eclipse, the apparent radius of the Sun and the Moon will be approximately equal. So, the approximate formula for overlapped area is obtained by assuming that the radius of both the circles to be equal.

$$OverlappedArea = \left(2rCos^{-1}\left(\frac{d}{2r}\right)\right) - d\sqrt{r^2 - \frac{d^2}{4}} \quad (6)$$

where r is the radius of both the circles.

### G. Distance between earth and the Moon

The apparent size of the Moon varies as a function of distance from the Earth. As the Moon moves away from the Earth, the apparent size of the Moon decreases and the size increases as the Moon approaches its perigee. From the images taken during eclipse, the apparent size of the Moon is obtained. This data is be used to calculate the approximate distance between the Earth and the Moon. The relationship between the Sun, the Earth and the Moon during an ideal solar eclipse is geometrically visualized as shown in figure 6.

In the image, the vertical line AB depicts the Sun and the line ED depicts the Moon. During an ideal solar eclipse, the apparent sizes of both the Sun and the Moon are equal and thus the entire Sun's disk will be obscured. This situation is used to obtain the distance between the Earth and the Moon. In the image, $x_1$ is the distance between the Earth and the Sun and $x_2$ is the distance between Earth and the Moon. The terms $y_1$ and $y_2$ are the actual radius of the Sun and the Moon respectively. In the image, let $r_1$ and $r_2$ are the lengths of the hypotenuse of the triangles ABC and EDC respectively.

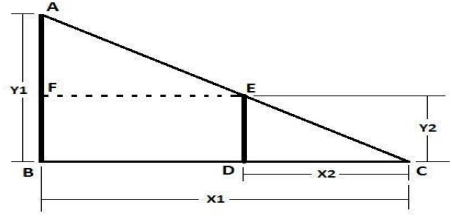

Figure 6. Relation between the Sun, the Moon and the observer during an ideal solar eclipse.

From the figure, the distance between the observer and the Moon is given by the formula

$$x_2 = \frac{x_1 y_2}{y_1} \quad (7)$$

The above equation can be applied only when the apparent size of the Moon and the Sun are equal. Since the size of the Moon keeps changing, the radius of the Sun, in the above equation, is replaced by the radius of the region on the surface of the Sun that is obscured by the Moon. The radius of the region obscured by the Moon, S_RadObs, is given by the formula

$$S\_RaObs = \frac{i\_m\_radius}{i\_s\_radius} * S\_ActualRadius \quad (8)$$

where i_m_radius is the radius of the Moon obtained from the image, i_s_radius is the radius of the Sun obtained from the image and S_ActualRadius is the actual radius of the Sun.

So, the formula to obtain the distance between the Earth and the Moon is

$$\frac{D\_Sun * m\_radius}{S\_RadObs} + radius\_earth \quad (9)$$

where D_Sun is actual distance between the Earth and the Sun, m_radius is the actual radius of the Moon and radius_earth is the actual radius of the Earth.

Since the observer is on the surface of the Earth, the distance between the Moon and the Earth is obtained by summing the radius of the earth and the value $x_2$.

Due to the nature of the orbit of Earth around the Sun, D_Sun is assumed to be a constant since the variation of Earth's orbital radius is negligible when compared to the variations in the lunar orbital radius.

## III. Conclusion

We have presented a new and efficient method to calculate various parameters regarding solar eclipse. The proposed algorithm not only calculates the appropriate parameters but also produces the results in a shorter period of time. The algorithm was simulated using MATLAB and when executed on an INTEL Pentium D processor at 2.66 GHz, the algorithm takes 6seconds to analyze an image with eclipse and 4 seconds to analyze an image without eclipse. .